\documentclass[final] {aipproc}
\layoutstyle{8x11double}

\usepackage{amsmath}      
\usepackage{amssymb}
\usepackage{amsthm}
\usepackage{array}
\usepackage{bbm}
\usepackage{color}
\usepackage{pgfsys}
\usepackage{epsfig}
\usepackage{latexsym}
\newcommand{\beqan}{\begin{eqnarray*}}
\newcommand{\eeqan}{\end{eqnarray*}}   
\newcommand{\ba}{\begin{array}}
\newcommand{\ea}{\end{array}}
\newcommand{\no}{\nonumber}

\newcommand\lsim{\mathrel{\rlap{\lower4pt\hbox{\hskip1pt$\sim$}}
    \raise1pt\hbox{$<$}}}
\newcommand\gsim{\mathrel{\rlap{\lower4pt\hbox{\hskip1pt$\sim$}}
    \raise1pt\hbox{$>$}}}

\newcommand{\lra}{\longrightarrow}

\def\slch#1{\setbox0=\hbox{$#1$}\dimen0=\wd0%
\setbox1=\hbox{/}\dimen1=\wd1%
\ifdim\dimen0>\dimen1%                     
\rlap{\hbox to
\dimen0{\hfil/\hfil}}#1\else                                     
\rlap{\hbox to \dimen1{\hfil$#1$\hfil}}/\fi}

\newcommand{\mrm}{\mathrm}

\begin{document}

\title{Chiral extrapolation of SU(3) amplitudes}


\author{Gerhard Ecker}{
  address={University of Vienna, Faculty of Physics, Boltzmanngasse 5,
  A-1090 Wien, Austria}
}

\begin{abstract}
Approximations of chiral $SU(3)$ amplitudes at NNLO are
proposed to facilitate the extrapolation of lattice data to the
physical meson masses. Inclusion of NNLO terms is essential for
investigating convergence properties of chiral $SU(3)$ and for
determining low-energy constants in a controllable fashion. The
approximations are tested with recent lattice data for the ratio of
decay constants $F_K/F_\pi$.
\end{abstract}

\maketitle

%%%%%%%%%%%%%%%%%%%%%%%%%%%%%%%%%%%%%%%%%%%%
%% MAINMATTER
%%%%%%%%%%%%%%%%%%%%%%%%%%%%%%%%%%%%%%%%%%%%

\section{Motivation}
Lattice QCD has made enormous progress in the light quark sector. Since
most lattice studies employ quark masses with $m_q^{\mrm{lattice}} >
m_q^{\mrm{phys}}$ an extrapolation in meson masses is needed. Although
chiral perturbation theory (CHPT) amplitudes \cite{CHPT} are available
to NNLO for most quantities of interest \cite{Bijnens:2006zp} the NNLO
contributions are usually quite involved. In addition, they are mostly
available only in numerical form for chiral $SU(3)$.

In the lattice community, the convergence properties of chiral $SU(3)$
are generally considered problematic. For a thorough discussion of
this issue, NLO amplitudes are not sufficient. Moreover, using quark
masses with $m_s^{\mrm{lattice}} < m_s^{\mrm{phys}}$ would be very
useful for assessing the convergence properties of chiral $SU(3)$
\cite{Bazavov:2009bb}. 

Tuning the quark masses offers a new environment for extracting chiral
low-energy constants (LECs). This is especially welcome for LECs that
are difficult to access phenomenologically,  both at NLO and at NNLO. 

For all these reasons, analytic approximations of NNLO CHPT amplitudes
are expected to be very useful. In this talk, I describe recent
attempts to derive such approximations \cite{Ecker:2010nc} that are
\begin{itemize} 
\item more sophisticated than the double-log approximation 
\cite{Bijnens:1998yu},
\item yet much simpler than the full numerical expressions. 
\end{itemize}

\section{CHPT at NNLO}
\begin{figure}
  \includegraphics[height=.3\textheight]{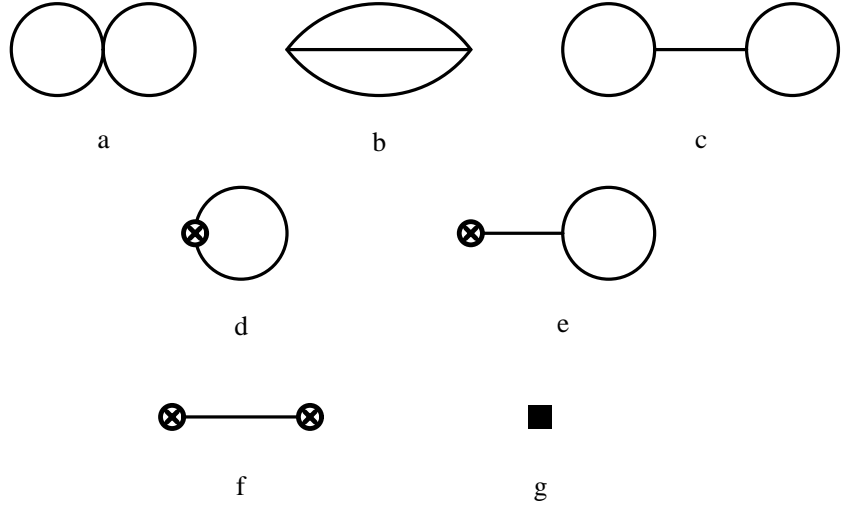}
\label{fig:genfunc}
  \caption{Skeleton diagrams for the generating functional
  $Z_6$ of  $O(p^6)$. Simple dots, crossed circles, black box denote
  vertices from LO, NLO, NNLO Lagrangians,
  respectively. Propagators and vertices carry the full tree structure 
  associated with the lowest-order Lagrangian.}
\end{figure}

A compact representation of CHPT is in terms of the generating
functional of Green functions \cite{CHPT}. The NNLO functional $Z_6$
of $O(p^6)$ is itself a sum of different contributions shown pictorially
in Fig.~\ref{fig:genfunc}. I concentrate here on the so-called
irreducible contributions \cite{Bijnens:1999hw,Ecker:2010nc} 
\begin{eqnarray}
Z_6^{\rm a+b+d+g} &=&  \int \!\! d^4x \left\{ \left[C_a^r(\mu)
  \right. \right.
  \\[.2cm] 
& \hspace*{-3.2cm} + &    \hspace*{-2cm} \left. \left. 
\displaystyle\frac{1}{4 F_0^2} \left(4\, \Gamma_a^{(1)} \,L  -
\Gamma_a^{(2)} \,L^2  + 2\, \Gamma_a^{(L)}(\mu) L
\right)\right]  O_a(x)  \right. \no \\[.2cm] 
& \hspace*{-3.2cm} + &   \hspace*{-2cm} \left.  
\displaystyle\frac{1}{(4\pi)^2} \left[L_i^r(\mu) - 
  \displaystyle\frac{\Gamma_i }{2} L \right] H_i(x;M) 
 +  \displaystyle\frac{1}{(4 \pi)^4}  K(x;M) \right\} \no~.
\end{eqnarray} 
%%%%%%%%%%%%%%%%%%%%%%%%%%%%%%%%%%%%%%%%%%%%
%% Sample figure:
%%
%% The option [height=...] scales the picture to the given height,
%% without it it would be printed at its nominal size
%%%%%%%%%%%%%%%%%%%%%%%%%%%%%%%%%%%%%%%%%%%%
\noindent The $C_a^r$ are LECs of $O(p^6)$, $O_a(x)$ the associated
monomials of the chiral Lagrangian \cite{Bijnens:1999sh}. The chiral
log $L$ is given by $L = 1/(4\pi)^2 \ln{M^2/\mu^2}$, $L_i^r$ are
LECs of $O(p^4)$ and $F_0$ is the pseudoscalar decay constant in the
chiral $SU(3)$ limit. The coefficients $ \Gamma_a^{(1)}$, $\Gamma_a^{(2)}$
and $\Gamma_a^{(L)}$ are listed in Ref.~\cite{Bijnens:1999hw}.
$H_i(x;M)$ (1-loop) and  $K(x;M)$ (2-loop) are nonlocal
functionals. $Z_6^{\rm a+b+d+g}$ and $Z_6^{\rm total}$ are independent
of both scales $\mu, M$.

\section{Approximations}
The drawback of the explicit form of the generating functional of
$O(p^6)$ for chiral $SU(3)$ is that 
the 2-loop contribution $K(x;M)$ is in general
only numerically available.  
The approximation suggested in Ref.~\cite{Ecker:2010nc} (referred to
as Approximation I from now on) has the following properties:
\begin{itemize} 
\item $ K(x;M)$ (and the reducible 2-loop contributions) are dropped;
\item All chiral logs are kept;
\item $Z_6^{\rm (appr. \,I)}$ is still invariant with respect to the
  renormalization scale $\mu$;
\item The residual dependence on the scale $M$ (appearing only in the
  chiral log $L$) is the only vestige of the 2-loop part;
\item Approximation I respects the large-$N_c$ hierarchy of $O(p^6)$
  contributions: \\
  \hspace*{1cm} $C_a, L_i L_j ~\lra ~L_i \times $\,loop $~\lra$  ~2-loop
\item Only tree and 1-loop amplitudes are needed;
\item Unlike the double-log approximation \cite{Bijnens:1998yu}, it
  allows for the extraction of LECs with the correct scale dependence.    
\end{itemize} 

One important question remains to be answered: how trustworthy is this
approximation? As the following examples will show, the answer depends
on the quantity under consideration.

\section{Applications}
\subsection{$F_K/F_\pi$}
The ratio of pseudoscalar decay constants $F_K/F_\pi$ exhibits a
picture-book type of chiral expansion as shown in Table
\ref{tab:physmass}. The separately  scale-dependent contributions of
$O(p^6)$ are given for $\mu=770$ MeV. The entries for ``numerical
results'' were provided by Bernard and Passemar
(Ref.~\cite{Bernard:2009ds} and private communication).   
\begin{table}
\begin{tabular}{ccccc}
\hline
  & \tablehead{1}{c}{b}{$O(p^4)$}
  & 
  & \tablehead{1}{c}{b}{$O(p^6)$}   \\
  & & 2-loop & $L_i \times$ loop & tree \\[.1cm]
numerical results \cite{Amoros:1999dp,Bernard:2009ds} & \hspace*{.2cm}
  0.14 \hspace*{.2cm}  & 0.002 & 0.051 & 
  0.008 \\[.1cm] 
Appr. I ($M=M_K$) &      & - 0.030 & & \\[.1cm]
Appr. II ($M=M_K$) &      & - 0.011 & & \\[.1cm]  
\hline
\end{tabular}
\caption{Chiral expansion of $F_K/F_\pi -1$} 
\label{tab:physmass}
\end{table}

For an $SU(3)$ quantity normalized to one at LO, successive terms in
the chiral expansion are expected to show the following generic
behaviour:  
\begin{center} 
\begin{tabular}{ccc}
\hspace*{.3cm} $O(p^4)$ \hspace*{.3cm} & $O(p^6)$ & $O(p^8)$  \\[.1cm] 
$\lsim 0.3$ & \hspace*{.1cm} $\lsim 0.3^2 = 0.09$ \hspace*{.1cm} &
$\lsim 0.3^3 = 0.027$ 
\end{tabular}  
\end{center}    
This suggests as criterion for an acceptable NNLO approximation that
the accuracy should not be worse than 3 $\%$, the typical size of a
term of $O(p^8)$. Table \ref{tab:physmass} shows that this criterion
is only barely met by Approximation I. A possible modification
consists in ignoring the strict large-$N_c$ counting to include also
products of 1-loop functions originating from diagrams a,c in
Fig.~\ref{fig:genfunc}. As Table \ref{tab:physmass} indicates, this
Approximation II \cite{EMNprep} satisfies the criterion for an
acceptable approximation.

In addition to two combinations of LECs of $O(p^6)$, also various 
$L_i$ appear in $F_K/F_\pi$. At $O(p^4)$, only $L_5$ enters. We
have therefore extracted $L_5$, $ C_{14} + C_{15}$ and $C_{15} + 2
C_{17}$ from a fit of the recent lattice data of the BMW Collaboration
\cite{Durr:2010hr}, using for the remaining $L_i$ (appearing only at
$O(p^6)$) the
values of fit 10 of Ref.~\cite{Amoros:1999dp}. The results
\cite{Ecker:2010nc,EMNprep} are displayed in Table \ref{tab:results}. 
\begin{table}
\begin{tabular}{ccccc}
\hline
  & \tablehead{1}{c}{b}{$F_K/F_\pi$}
  & \tablehead{1}{c}{b}{$10^3 L_5^r$}
  & \tablehead{1}{c}{b}{$10^3 (C_{14}^r + C_{15}^r) ~{\rm GeV}^2$}
  & \tablehead{1}{c}{b}{$10^3 (C_{15}^r + 2 C_{17}^r) ~{\rm GeV}^2$}
  \\[.25cm] 
Appr. I ($M=M_K$) &  $1.198(5)$    & $0.76(9)$ & $0.37(8)$ & $1.29(16)$ 
 \\[.1cm] 
Appr. II ($M=M_K$) & $1.200(5)$    & $0.75(9)$ & $0.20(7)$ & $0.71(15)$ 
 \\[.1cm]
BMW \cite{Durr:2010hr} & \hspace*{.2cm} $1.192(7)_{\rm stat}(6)_{\rm syst}$
  \hspace*{.2cm}  & & & \\[.1cm] 
\hline
\end{tabular}
\caption{Fit results for $F_K/F_\pi$ and LECs (statistical errors
  only) }
\label{tab:results}
\end{table}
The fitted values of $F_K/F_\pi$ agree with the detailed analysis of
Ref.~\cite{Durr:2010hr}. For both $F_K/F_\pi$ and $L_5$, there is
practically no difference between the two approximations but the LECs
of $O(p^6)$ show a bigger spread. The results
are shown for $M=M_K$. For a discussion of the $M$-dependence,
I refer to Refs.~\cite{Ecker:2010nc,EMNprep}. The fit also
demonstrates that NNLO terms are absolutely essential. While the NNLO
fit (Approximation II) is well behaved ($\chi^2$/dof = 1.2, statistical errors
only), the NLO fit with the single parameter $L_5$ is a catastrophe
($\chi^2$/dof = 4).  

\subsection{Other applications}
\paragraph{$F_\pi/F_0$ in chiral $SU(3)$}
From the chiral point of view, the interest is not so much in
``deriving'' $F_\pi$ from lattice data but rather to extract the still
badly known $F_0$ and the NLO LEC $L_4$. $F_0$ sets the scale of the
chiral $SU(3)$ expansion and is therefore essential for an appraisal
of convergence properties. The comparison of Approximation I with
available numerical results (Ref.~\cite{Bernard:2009ds} and
private communication) is very encouraging in this case
\cite{EMNprep}. 

\paragraph{$K_{l3}$ vector form factor $f_+(t)$}
A crucial quantity for a precision determination of the CKM element
$V_{us}$ is the value of $f_+(0)$. For at least two reasons
\cite{EMNprep}, both approximations discussed here do not appear very
promising in this case.
\begin{itemize} 
\item The chiral expansion of $f_+(0)$ shows a rather atypical
  behaviour. 
\item The approximations considered so far do not match with
  available numerical results
  \cite{Post:2001si,Bijnens:2003uy,Bernard:2009ds}.  
\end{itemize}

\section{Conclusions}
\begin{enumerate} 
%\begin{itemize} 
\item We have provided user-friendly extrapolation formulas
  \cite{Ecker:2010nc,EMNprep} for $N_f=2+1$ lattice data.
\item The proposed approximations are superior to NLO and chiral
  double-log approximations and should be useful for 
  investigating the convergence properties of chiral $SU(3)$.
\item They allow for the extraction of NLO and NNLO LECs with the
  correct scale dependence.
\item Analysis of the BMW data \cite{Durr:2010hr} for $F_K/F_\pi$
  provides the best available determination of $L_5$, assuming the
  large-$N_c$ hierarchy $|L_4|,|L_6| \ll L_5$.
\item The proposed approximations are intended to be used by
  lattice groups: they offer considerably more insight than, e.g.,
  polynomial fits.  
%\end{itemize} 
\end{enumerate} 

\begin{theacknowledgments}
The results presented here have grown out of a very pleasant
collaboration with Pere Masjuan and Helmut Neufeld. I also thank
V\'eronique Bernard, Steve Gottlieb and Emilie Passemar for helpful
discussions.  
\end{theacknowledgments}


\begin{thebibliography}{10}
\bibitem{CHPT}
  J.~Gasser and H.~Leutwyler,
  %``Chiral Perturbation Theory To One Loop,''
  Annals Phys.\  {\bf 158} (1984) 142; Nucl.\ Phys.\  B {\bf 250}
  (1985) 465. 
  %%CITATION = APNYA,158,142;%%

\bibitem{Bijnens:2006zp}
  J.~Bijnens,
  %``Chiral Perturbation Theory Beyond One Loop,''
  Prog.\ Part.\ Nucl.\ Phys.\  {\bf 58} (2007) 521
  [arXiv:hep-ph/0604043].
  %%CITATION = PPNPD,58,521;%%

\bibitem{Bazavov:2009bb}
  A.~Bazavov, C.~Bernard, C.~E.~DeTar {\it et al.},
  %``Nonperturbative QCD simulations with 2+1 flavors of improved
  %staggered quarks,'' 
  Rev.\ Mod.\ Phys.\  {\bf 82 } (2010)  1349.
  [arXiv:0903.3598 [hep-lat]].

\bibitem{Ecker:2010nc}
  G.~Ecker, P.~Masjuan and H.~Neufeld,
  %``Chiral extrapolation and determination of low-energy constants
  %from lattice data,'' 
  Phys.\ Lett.\  B {\bf 692 } (2010)  184.
  [arXiv:1004.3422 [hep-ph]]
.

\bibitem{Bijnens:1998yu}
  J.~Bijnens, G.~Colangelo and G.~Ecker,
  %``Double chiral logs,''
  Phys.\ Lett.\  B {\bf 441} (1998) 437
  [arXiv:hep-ph/9808421].
  %%CITATION = PHLTA,B441,437;%%

\bibitem{Bijnens:1999hw}
  J.~Bijnens, G.~Colangelo and G.~Ecker,
  %``Renormalization of chiral perturbation theory to order p**6,''
  Annals Phys.\  {\bf 280} (2000) 100
  [arXiv:hep-ph/9907333].
  %%CITATION = APNYA,280,100;%%

\bibitem{Bijnens:1999sh}
  J.~Bijnens, G.~Colangelo and G.~Ecker,
  %``The mesonic chiral Lagrangian of order p**6,''
  JHEP {\bf 9902} (1999) 020
  [arXiv:hep-ph/9902437].
  %%CITATION = JHEPA,9902,020;%%


\bibitem{Bernard:2009ds}
  V.~Bernard and E.~Passemar,
  %``Chiral Extrapolation of the Strangeness Changing K \pi Form Factor,''
  JHEP {\bf 1004} (2010) 001
  [arXiv:0912.3792 [hep-ph]].
  %%CITATION = JHEPA,1004,001;%%

\bibitem{EMNprep}
G.~Ecker, P.~Masjuan and H.~Neufeld, in preparation.

\bibitem{Durr:2010hr}
  S.~D\"urr {\it et al.},
  %``The ratio FK/Fpi in QCD,''
  Phys.\ Rev.\  D {\bf 81} (2010) 054507
  [arXiv:1001.4692 [hep-lat]].
  %%CITATION = PHRVA,D81,054507;%%

\bibitem{Amoros:1999dp}
  G.~Amor\'os, J.~Bijnens and P.~Talavera,
  %``Two-point functions at two loops in three flavour chiral perturbation
  %theory,''
  Nucl.\ Phys.\  B {\bf 568} (2000) 319
  [arXiv:hep-ph/9907264].
  %%CITATION = NUPHA,B568,319;%%


\bibitem{Post:2001si}
  P.~Post and K.~Schilcher,
  %``K(l3) form-factors at order p**6 of chiral perturbation theory,''
  Eur.\ Phys.\ J.\  C {\bf 25 } (2002)  427.
  [hep-ph/0112352].

\bibitem{Bijnens:2003uy}
  J.~Bijnens and P.~Talavera,
  %``K(l3) decays in chiral perturbation theory,''
  Nucl.\ Phys.\  B {\bf 669 } (2003)  341.
  [hep-ph/0303103].


\end{thebibliography}
\end{document}